# Assessment of Physical Properties of Water Repellent Soils

Mahta Movasat[1] and Ingrid Tomac[2], Ph.D., M. ASCE

[1]Structural Engineering Department, University of California San Diego, 9500 Gilman Drive, La Jolla, CA 92093-0085, email: mmovasat@ucsd.edu; corresponding author

[2]Structural Engineering Department, University of California San Diego, 9500 Gilman Drive, La Jolla, CA 92093-0085, email: itomac@ucsd.edu

**Abstract**

This note presents a comprehensive characterization of physical and mechanical properties of water repellent (hydrophobic) soil collected from Cleveland National Forest in California immediately after the Holy Fire, 2018, and delineates comparisons with chemically induced hydrophobic sand in the laboratory. Hydrophobicity is a particle surface characteristic that governs different levels of attraction between water molecules and solid particles. Wildfires can cause different levels of hydrophobicity in shallow soil layers based on fire severity, vegetation, and chemical structure of the soil. Natural and chemically induced regular and hydrophobic sands are characterized by grain size distribution, water retention curve, water contact angle and electron microscopic imaging, including the relationship between water entry value and the drop contact angle in hydrophobic soil. Comparative knowledge of natural and chemically induced hydrophobic soil properties will help future research to better predict soil behavior and improve insights into post-wildfire soil erosion and mudflow mechanisms. This note contributes to a database of wildfire-induced hydrophobic soil with detailed properties and assesses the applicability of laboratory made hydrophobic soils for studying mudflows by comparison to the natural water repellent soil collected from the burned site.

**Introduction**

During wildfires, accumulated organic matter such as plant root exudates, certain fungal species and surface wax from plant leaves decomposes and volatilizes into soil and induces soil water repellency, i.e. hydrophobicity, in shallow soil layers (Doerr et al. 2000). A mechanism that underlies hydrophobicity is a weak attraction between the molecules of liquid water and solid due to low-energy soil grain surface created by the deposited organic matter (Leelamanie et al. 2008). Hydrophobicity of surficial soil layers causes rain soil erosion, overland flow, rain-splash detachment and preferential flow paths in burned areas (Debano 2000, Scott and Van Wyk 1990). Runoff-dominated processes have been widely reported in recently burned areas (Cannon et al. 2001, Nyman et al. 2011). Previous studies looked at the geographical patterns of the runoff and have shown that grain size distribution, shape, and characteristics of overland flow affect the erosion mechanism and sediment transportation (Asadi et al. 2011, Rose et al. 2007). However, the relevant soil properties, such as are the soil hydrophobicity levels and infiltration characteristics have not yet been extensively studied and related to runoff. The relationship between the volumetric water content ($\theta$) and the matric suction ($\psi$) is called soil water retention curve (SWRC) and determines the hydraulic conductivity, drainage, solute movement and suction distribution in soil (Kern 2010, Marmur 1992), and can be fitted to the van Genuchten empirical model (Leong and Rahardjo, 1997). Adli et al. (2014) showed that the air entry value which is a controlling parameter in SWRC plays an important role in failure time of slopes during rainfall.

Dynamics of droplet shape and movement on surface can improve understanding the mechanism of erosion and onset of mudflows. Surface energy and wettability are two factors that affect the behavior of water droplet on the surface (Boinovich and Emelyanenko 2008). Soil minerals with higher density of charge and polar groups have a higher wettability and affinity for water (Bachmann et al. 2007). However, also smaller values of interfacial tension have been found for minerals in soil by Miyamoto (1971). Wetting characteristics of soil demonstrates the spreading ability of water on the surface and can indicate the interfacial forces between soil, liquid and gas, which can be measured by soil water contact angle. Water contact angles are classified as ultra-hydrophilic (≈0º), hydrophilic (0º-90º), hydrophobic (90º-150º) and ultra-hydrophobic (>150º) (Chieng et al. 2019). In hydrophobic surfaces, the contact angle is low and thus the droplet tends to roll down and move easily (Ragesh et al. 2014).

This study is motivated by a necessity of filling in the limited amount of available data of natural hydrophobic post-wildfire soil properties, which are difficult to collect and not frequently extensively studied. Measurements of the water drop penetration time, the contact angle, the direct shear strength, and the soil water characteristics are conducted on hydrophobic and regular sand. Scanning Electron Microscope (SEM) and optical microscope images are taken from both samples to investigate the micro-properties of the hydrophobic soil. Relating the drop contact angle of hydrophobic soil to the water entry value from SWRC can help assess slope stability. Knowledge of soil characteristics will aid in better understanding the mechanisms of erosion and debris flow in hydrophobic soils, which has a potential to reduce life hazards and enhance early warning systems.

**Materials and Methodology**

In August 2018, the Holy Fire burned 23,000 acres of Cleveland National Forest in California. Hydrophobicity was reported by Burned Area Emergency Response (BAER) team in almost all burned areas with about 71% burned with moderate severity and chaparral dominated vegetation (Nicita and Halverson 2018). The precipitation in area ranges from 13 in to 23 in per year, with 78% of storms happening between October and April, including occasional summer thunderstorms. Geologically young and steep mountains in this area result in higher risk of a debris flow, rock fall and slumping especially in the severely burned steep areas (Nicita and Halverson 2018). We have collected hydrophobic soil samples directly from the area of Cleveland national forest in October 2018. After removing the ash layer from the soil surface, we performed water droplets beading assessment and tested in-situ hydrophobicity. Then, we carefully obtained soil samples from approximately 4 cm thick hydrophobic soil layer, sealed them in airtight container and removed remaining roots in the laboratory by sieving. Ottawa F-65 sand, a rounded-particle soil, is another batch of tested soil which is processed to become artificially hydrophobic. After drying for 24 hours, the Ottawa F-65 sand is submerged in 10% by volume $V_{\text{n-octyltriethoxysilane}}/V_{\text{isopropyl-alcohol}}$ solution for the next 48 hours and the sand is washed to remove any reactive compound and oven-dried for 24 hours (Karim et al. 2018). Hydrophobicity level is determined by water drop penetration time test (WDPT) for all soil samples. The WDPT places 50 μL water droplets on the soil surface and measures the time required for the complete penetration. The water repellency is categorized with the static water contact angle is measurement by sessile droplet method (Lee et al. 2015). A double-sided adhesive tape is fixed on a microscopic glass, oven-dried soil is sprinkled on the 2 cm by 2 cm tape and tapped carefully to remove any excess soil. A 100-gr weight is used to press the soil particles to the tape and the procedure is repeated two times for all the samples. Five drops of deionized water are placed on the soil layer by a pipette, where each drop's volume is 1.7 mm$^3$. A scientific Phantom Miro C320 camera images the droplet with high precision 640x480 pixels immediately, and after 5 and 10 min. The SWRC is measured using the hanging column method (ASTM D6836-02) for determining the wetting-path curve, since samples might lose the hydrophobicity once saturated. A dry soil sample residing in the retaining ring with filter-paper bottom is twisted for 45º when putting to a glass funnel to ensure the contact between the sample and a saturated (for 24h) porous plate. A thin plastic film covers the funnel to avoid evaporation. Lowering the funnel from the balance point applies various suctions to the specimen, while entered or expelled water level is measured by a capillary tube. Water elevation is recorded in each step and water retention curves are obtained. Direct shear tests are conducted on dry specimens subjected to three different vertical stresses of $\sigma_1$=83 kPa, $\sigma_2$=134 kPa, and $\sigma_3$=180 kPa in displacement-control mode with a shearing rate of 0.5 mm/min. All specimens are tamped in three identical dry layers in a 63.5x63.5 mm square-shaped box and sheared to failure. To obtain Scanning Electron Microscope (SEM) images of the site soil and treated and untreated sand, the soil sample is coated in a sputter coater device (Emitech K575X) with iridium to provide a conductive surface to scan the sample and images are taken (FEI Quanta FEG 250 device). In addition to SEM images, energy-dispersive *X*-ray microanalysis (EDX) is conducted for each sample to map chemical elements distribution and characterize constitutive soil minerals present in the sample.

**Results**

Soil particle distribution for both the Cleveland and Ottawa F-65 sands is shown in Fig. 1. The coefficient of uniformity and curvature are 4.7 and 0.8 respectively for Cleveland soil, which is classified as poorly graded sand (SP). The grain distribution of Ottawa F-65 indicates that the sand is also SP, with a coefficient of uniformity and curvature of 1.61 and 0.96. WDPT test results show that the sample taken from Cleveland national forest and the hydrophobic Ottawa F-65 sand are strongly and severely water repellent respectively (Table 1).

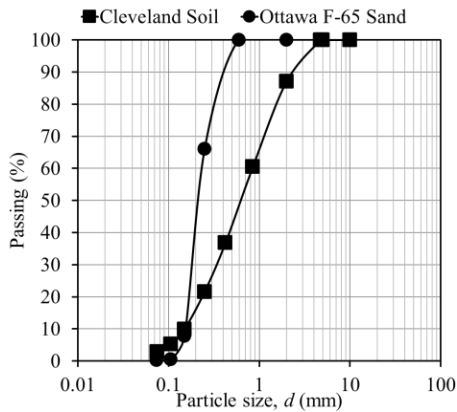

**Fig. 1.** Grain size distribution of Cleveland hydrophobic soil and Ottawa F-65Sand

**Table 1.** Water drop penetration time test results

| Soil | WDPT | Classification |
|---|---|---|
| Cleveland | 100 s | Strongly water repellent |
| Ottawa F-65 regular | 5 s | Wettable |
| Ottawa F-65 hydrophobic | 3600 s | Severely water repellent |

Water drop contact angle of both hydrophobic and regular Ottawa F-65 sand is measured using sessile drop method (Fig. 2). The photos of the drop are taken immediately after the drop was placed and after 10 s, 1 min and 10 min. The images show that the initial contact angle in hydrophobic sample is 115° and 61° in regular sample, the angle has changed over time in both samples, but the rate of change in regular sample is more significant than in hydrophobic.

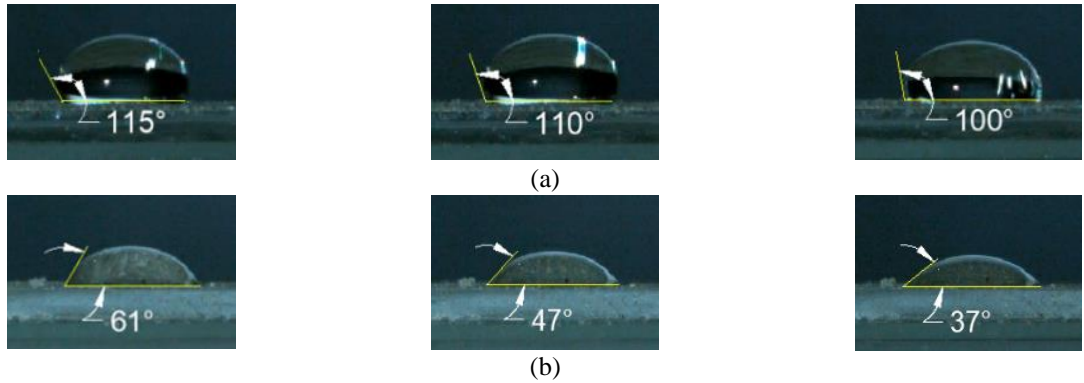

(a)

(b)

**Fig. 2.** Contact angle α measurements in (a) hydrophobic and (b) regular sand after putting the drop, 5 minutes and 10 minutes later from left to right

The SWRCs for the Cleveland, Ottawa F-65 regular and hydrophobic sand with a unit weight of 1.66 g/cm$^3$ are shown in Fig. 3. Independently fitted parameters, namely α and *N*, in van Genuchten SWRC equation (Eq. 1) are shown for all materials in Fig. 3:

$$\theta(\psi) = \theta_r + \frac{\theta_s - \theta_r}{[1 + (\alpha|\psi|)]^{1-1/N}} \qquad (1)$$

where $\theta(\psi)$ is the water retention curve, $|\psi|$ is the suction pressure, $\theta_s$ is the saturated water content, $\theta_r$ is the residual water content, α is the parameter related to the inverse of air entry suction and *N* is the measure of the pore size distribution. The results of hydrophobic and regular Ottawa F-65 sand indicate that the final volumetric water content of hydrophobic soil has reduced about 7 times. The van Genuchten fitted curve also indicates that the air entry value has decreased for the hydrophobic sample. The final water content of Cleveland and Ottawa F-65 hydrophobic specimens is about 7.4% and 5% respectively when reaching zero suction. The SWRC results clearly indicate and confirm that hydrophobic soils resist water infiltration and absorb less water than regular soil during rainfall event. The Cleveland sample is also re-tested after one year and the results indicate an increase in water content from 7.4% to 19.0% and a reduction of hydrophobicity effect on this sample. Lower water entry values, water content and the reduction of stored water were recorded in hydrophobic soil compared to regular soil. Results indicate decrease in mobility of water through pores between dry hydrophobic surfaces initially filled with air. Lower water entry values that are obtained for hydrophobic samples help better understand mechanisms of rainwater overflow, erosion and subsequent mudflows observed in burned slopes.

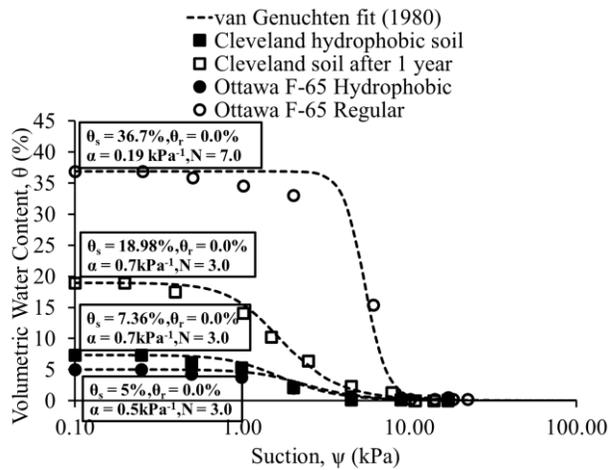 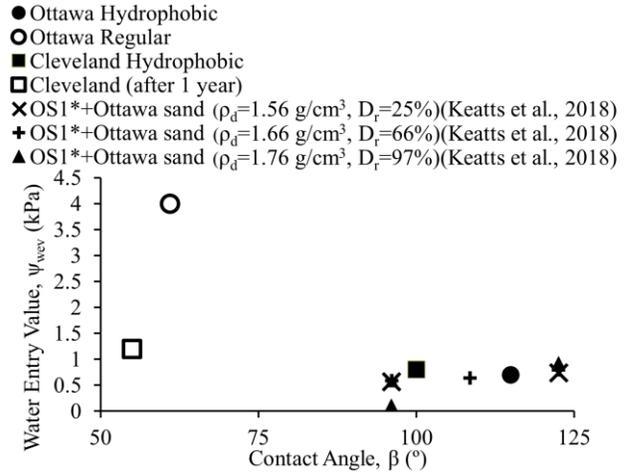

**Fig. 3.** Wetting curves from SWRC test

**Fig. 4.** Water entry value versus contact angle β, OS1*=organo-silane treated soil for hydrophobicity

Water entry values, obtained from the SWRC curves are shown in Fig. 4. Additionally the results are compared with the results of Organo-silane treated Ottawa sand from Keatts et al. (2018). The water entry value of hydrophobic samples decreases to approximately 1 kPa and below. Although the Cleveland soil lost its hydrophobicity after one year shown by the water contact angle drop to 55º (Fig. 5), the water entry value increases only slightly from 0.8 kPa to 1.2 kPa. The lower water entry value indicates the higher risk of failure and erosion in water repellent slopes. However, the relation of water entry value with contact angle for a recovered hydrophobic soil indicates that the risk of failure might be still high in recovered sites.

Direct shear test results Cleveland hydrophobic soil at a shearing rate of 0.5 mm/min and a unit weight of 15.5 kN/m$^3$, show a friction angle of 39°. Dry Ottawa F-65 regular and hydrophobic sand direct shear tests indicate that the friction angle decreases after the soil becomes hydrophobic. Both samples are tamped in three-layer and with a unit weight of 17.5 kN/m$^3$. The friction angle for Ottawa F-65 sand is 37° while it drops to 30° in hydrophobic Ottawa F-65 sand (Fig. 6). Similar trend is shown in a previous study by Karim et al. (2018).

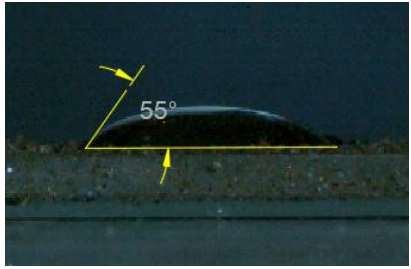

**Fig. 5.** Water contact angle of Cleveland soil measured 1 year after collecting the sample

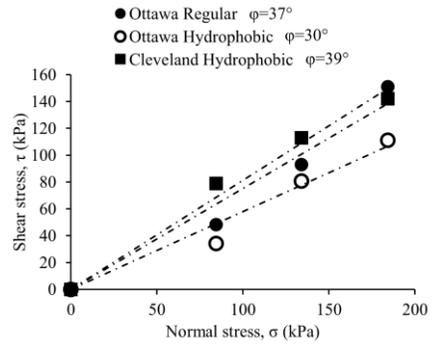

**Fig. 6.** Direct shear test results

Scanning electron microscopic image and EDX analysis are shown in the Fig. 7. a, b. The soil particles have angular shapes and their surface area is covered with minuscule flakes. Oxygen (O), Silicon (Si) and Aluminum (Al) constitute 55%, 18% and 10% of the atomic weight of the soil sample. Iron, Potassium, Magnesium, Titanium and Sodium are the other minor minerals found in the soil sample. Iridium is found in the soil since the sample was Iridium-coated. The minerals in soil indicate that the geological structure of the area can be consisted of quartz and alkali feldspars.

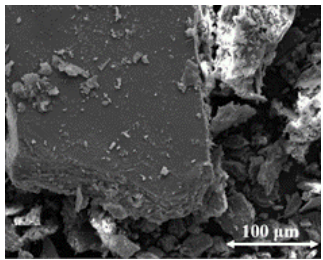
(a)
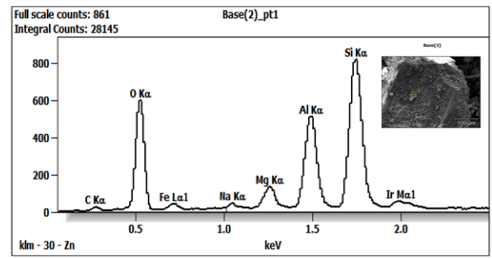
(b)

**Fig. 7.** (a) SEM image of Cleveland hydrophobic soil with a magnification of 100 μm, (b) EDX analysis

Fig. 8a and b show regular and hydrophobic Ottawa F-65 sand. The soil grains have rounded shape, and the regular sand surface is smother than hydrophobic sand surface. The hydrophobic sand surface has more irregularities and is coated with a thin layer of flakes in some parts. EDX analysis (Fig. 9. ) shows that the majority of the chemical composition of the soil is mainly Oxygen (O) and Silicone (Si) with an increased amount of Oxygen in hydrophobic sample due to the n-octyltriethoxysilane ($C_{14}H_{32}O_3Si$) coating.

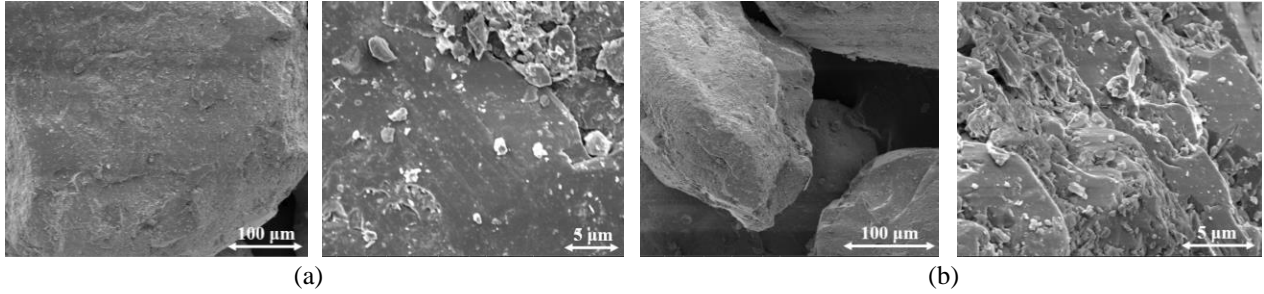

(a)                (b)

**Fig. 8.** SEM images of (a) Ottawa F-65 regular and (b) Ottawa F-65 hydrophobic sand with magnification of 100 μm and 5 μm from left to right

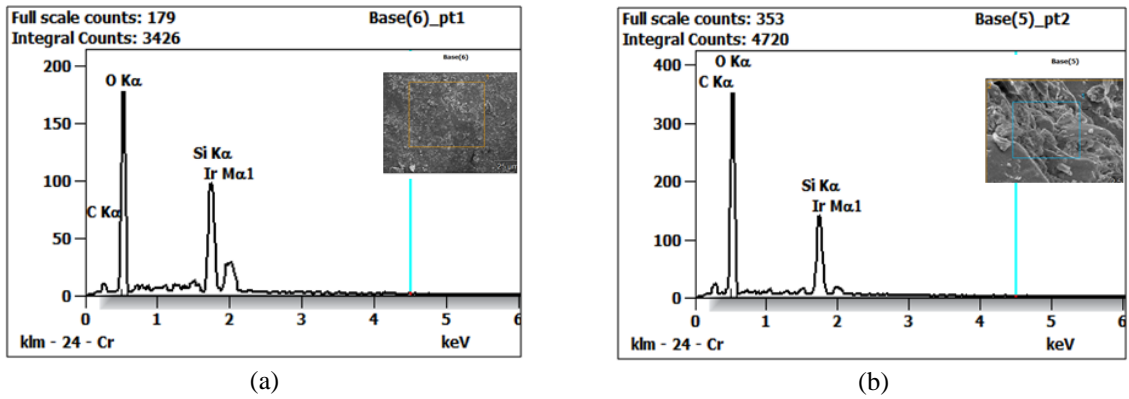

(a)                (b)

**Fig. 9.** EDX analysis of (a) Ottawa F-65 regular (b) Ottawa F-65 hydrophobic sand

**Conclusions**

Hydrophobicity has a significant effect on the hydraulic and physical characteristics of soil. Conversion in soil characteristics will lead to a change in soil behavior, which is critical during rainfall and flood in slopes. In this study, natural fire-induced hydrophobic soil is collected and tested from Cleveland National forest and laboratory-made hydrophobic soil is tested and compared to a regular soil of its own kind.

The results indicate a significant decrease in water retention in both fire-induced and laboratory-made hydrophobic soil samples. In Cleveland and Ottawa F-65 hydrophobic samples the final water content was about 7% and 5%. The decrease in water entry value for the hydrophobic sample was also measured, which has been previously found to indicate the higher risk of failure in slopes. The relation of water contact angle and water entry level can be a good indicator of the severity of water overflow in hydrophobic soils. In addition, the results from testing the Cleveland soil after one year shows a great reduction in hydrophobicity and increase in water content from 7% to 19%. Water contact angle results reveal the larger contact angle for Ottawa F-65 hydrophobic sample, with the reduction of angle during time is less significant for the hydrophobic sample.

The direct shear test results for laboratory-made hydrophobic sand demonstrated a decrease in friction angle from 37º in normal sand to 30º for the hydrophobic sample, which causes a reduction of shear strength in hydrophobic sample and will therefore increase the risk of failure in hydrophobic slopes. Scanning electron microscopic images illustrates a smoother surface for regular surface, while the hydrophobic sample has more irregularities on surface and a thin layer of flakes cover the surface. EDX analysis for each soil displayed that the Cleveland soil contained more minerals. However, the laboratory-made hydrophobic Ottawa F-65 sand was mainly consisted of Silica and Oxygen with higher amount of Oxygen in hydrophobic sample.

**Data Availability Statement**

All data, models, and code generated or used during the study appear in the submitted article.

**Acknowledgment**

Financial support of University of California, San Diego (UCSD) and Hellman Fellowship Foundation are greatly acknowledged. The opinions expressed in this paper are those of authors and not of UCSD or Hellman Foundation.

**Notation**

The following symbols are used in this paper:

$D_r$ = relative density;

$N$ = measure of the pore size distribution;

$\alpha$ = parameter related to the inverse of air entry suction;

$\beta$ = water contact angle;

$\theta_r$ = residual water content;

$\theta_s$ = saturated water content;

$\theta(\psi)$ = water retention curve;

$\rho_d$ = dry density;

$\sigma$ = normal stress;

$\tau$ = shear stress;

$\varphi$ = friction angle;

$|\psi|$ = suction pressure;

$\psi_{wev}$ = water entry value.